\documentclass{ar-1col-S2O}
\usepackage[numbers]{natbib}
\usepackage{url}
\usepackage{verbatim,epsfig,amsmath,amssymb,bm,epsf,graphicx,psfrag,bbold,amsthm,amsfonts}
\usepackage[bottom]{footmisc}
\usepackage{hyperref}
\usepackage{cleveref} 
\usepackage{enumitem}
\usepackage{framed}
\usepackage{mathrsfs}
\usepackage{esint}
\setlist{nosep}
\usepackage{placeins}
\usepackage[all]{xy}
\usepackage{color}
\usepackage[utf8]{inputenc}
\usepackage{float}
\usepackage{natbib}
\usepackage{tikz-cd}
\usepackage{verbatim}
\usepackage{leftidx}
\usepackage{physics}
\usepackage{makecell}

\usepackage{pifont}
\usepackage{tabularx}
\usepackage{ulem}
\usepackage{xcolor}

\newcommand{\bscco}{BSCCO }

\jname{Xxxx. Xxx. Xxx. Xxx.}
\jvol{AA}
\jyear{YYYY}
\doi{10.1146/((please add article doi))}

\begin{document}

\markboth{Pixley and Volkov}{Twisted Nodal Superconductors}

\title{Twisted Nodal Superconductors}

\author{J. H. Pixley$^{1,2}$ and Pavel A. Volkov$^3$
\affil{$^1$Department of Physics and Astronomy, Center for Materials Theory, Rutgers University, Piscataway, New Jersey 08854, USA; email: jed.pixley@rutgers.physics.edu}
\affil{$^2$Center for Computational Quantum Physics, Flatiron Institute, New York, New York 10010, USA}
\affil{$^3$Department of Physics, University of Connecticut, Storrs, Connecticut 06269, USA}}

\begin{abstract}
Recent proposals for the realization of time-reversal symmetry breaking and topological superconductivity in twisted nodal superconductors have led to a surge of theoretical and experimental studies of these systems, marking one of the newest entries in the rapidly growing field of moir\'e materials. The interplay between order parameters of the separate layers makes twisted superconductors unique, leading to additional emergent phenomena in regimes usually not of importance in  moir\'e physics, such as bulk interfaces and large twist angles.
We review the physics of twisted nodal superconductors, highlighting both similarities and qualitative differences with other moir\'e platforms. While inspired by the rise of moir\'e materials, the field is anchored in studies of unconventional superconductivity preceding the moir\'e era, which we discuss in detail. In addition to summarizing the developments at the present stage, we present a detailed outlook on the major open questions in the field 
and some of the most exciting future directions.
\end{abstract}

\maketitle

\tableofcontents

\section{Introduction}
\label{sec:intro}
The rise of two-dimensional (2D) van der Waals heterostructures~\cite{geim2013van,liu2016van} opened new horizons in the  control over solid state systems~\cite{andrei2021marvels}. The remarkable discoveries of emergent phenomena in twisted bilayer graphene (TBG)~\cite{li2010observation,andrei2020graphene} and  transition metal dichalcogenides (TMDs)~\cite{mak2022semiconductor} (summarized in Table \ref{tab1}) established a controllable way to drive weakly interacting semiconducting layers into a plethora of strongly correlated and topological quantum phases. 
Here, we review the progress in applying these principles to a different material class: nodal superconductors. Some of the first theoretical \cite{vishwanath2021proposals,can2021high,VolkovPRL-2023} and experimental \cite{glazman2022jc,lee2021twisted,zhao2023} results on twisted nodal superconductors (TNS) appeared around 2021, and the field has been rapidly expanding ever since.

\begin{textbox}[b!]\section{Defining Twisted Superconductors}
This review is on twisting {\it superconductors}, i.e. creating systems from two separate superconductors. This should be contrasted with materials that superconduct {\it only} when a twisted bilayer is formed, that we refer to as ``moir\'e superconductors''  e.g. twisted bilayer graphene~\cite{cao2018unconventional}, which is made out of non-superconducting graphene layers.
\label{textbox}
\end{textbox}

\begin{figure}[h]
\includegraphics[width=\linewidth]{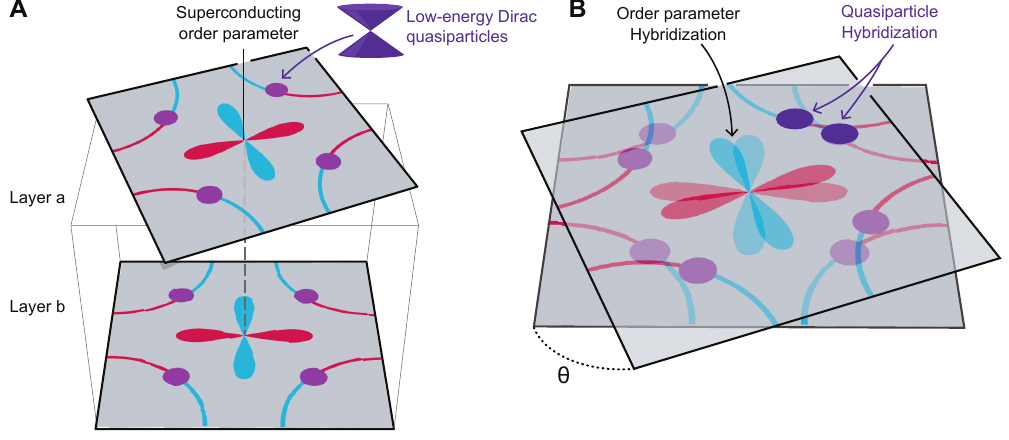}
\caption{{\bf Twisted nodal superconductors exemplified by a $d$-wave bilayer}. Squares denote the superconducting layers stacked on top of one another with a twist. {\it Left:} 
Each layer is characterized by two main degrees of freedom: the superconducting order parameter (center with a $d_{x^2-y^2}$ order parameter with positive/negative regions marked with blue/red) and low-energy Bogoliubov-de-Gennes quasiparticles with an anisotropic Dirac dispersion in momentum space (purple circles and cones). In individual layers, the Dirac points are positioned on the Fermi surface (blue/red colored lines) where the order parameter value changes sign (blue/red denotes the regime where the order parameter is positive/negative). {\it Right:} The coupling between two twisted nodal SC layers leads to interference effects that can modify both the state of order parameters (see Sec. \ref{sec:OP}) and Dirac BdG quasiparticles (Sec. \ref{sec:WF}). Each Dirac node defines a ``valley'' that contains two Dirac points (one from each layer).
}
\label{fig1}
\end{figure}

The defining feature of a superconductor is the presence of  long-range (or quasi-long range, in 2D materials) off-diagonal order, described by the complex order parameter $\Delta({\bf k})$, ${\bf k}$ being the crystalline momentum \cite{tinkham2004introduction,SigristUeda} (Fig. \ref{fig1}).
Conventional $s$-wave superconductivity, driven by  electron-phonon interactions\cite{BCS-PhysRev.108.1175-1957,mahan2013many}, is typically characterized by a uniform order parameter $\Delta({\bf k}) \approx \text{const}$. Unconventional superconductors \cite{stewart2017unconventional}, on the other hand, comprise a large class of material families, including  cuprates \cite{keimer2015,proust2019}, Fe-based \cite{stewart2017unconventional}, organics \cite{stewart2017unconventional}, heavy-fermions \cite{stewart2017unconventional}, and nickelates \cite{nickelates_rev}, where both the form of $\Delta({\bf k})$ and the mechanism behind it remain under present investigation. Theoretically, they can be classified by the symmetry of $\Delta({\bf k})$, that can be lower than that of the underlying crystalline lattice \cite{SigristUeda}. The existence of some non $s$-wave states is believed to have been established;  e.g., the $d$-wave state $\Delta({\bf k}) \propto \Delta_0(k_x^2-k_y^2)$ in cuprates \cite{Tsuei-RevModPhys.72.969-2000} and heavy fermions~\cite{allan2013imaging,zhou2013visualizing}. Other states, such as time reversal symmetry breaking~\cite{andersen2024spontaneous} ones, have a limited number of candidates and experimental evidence. 

Yet broader possibilities open, when considering the  structure of the excitation spectrum of superconductors, which are composed of the Bogoliubov-de-Gennes (BdG) bands of chargeless~\cite{kivelsoncharge} quasiparticles. The order parameter opens a spectral gap at the Fermi surface $\propto |\Delta({\bf k})|$. In the conventional case, the BdG bands are thus gapped similarly to a band insulator. In 2D {\it nodal} superconductors (such as the cuprates), in contrast, $\Delta({\bf k})$ vanishes along certain lines in momentum space, leading to the formation of gapless points, where they cross the Fermi surface (Fig. \ref{fig1}, purple disks). The BdG quasiparticle dispersion around such points is linear (but not isotropic) in momentum, i.e. it's a BdG Dirac cone (see Fig.~\ref{fig1} {\bf A} purple cone), calling for analogies with graphene (Table \ref{tab1}). Moreover, not all gapped superconductors are equal: unconventional gapped superconductors in 2D can have topological bands \cite{bernevig2013topological,sato2017topological} belonging to C, D, and DIII Altland-Zirnbauer classes~\cite{Schnyder-Ryu-2008,ryu2010topological}. This leads to the presence of gapless BdG edge modes on their boundary and within vortices~\cite{bernevig2013topological}. In some cases, these modes are Majorana fermions, holding potential for fault tolerant quantum computation~\cite{RevModPhys.80.1083-Nayak-2008,Mong-PhysRevX.4.011036-2014}. However, despite many proposed topological superconductor candidates, the definite experimental evidence for this state is lacking.

\begin{table}[h]
\tabcolsep3.5pt
\caption{Comparison of TNSs (focusing on $d$-wave spin singlet pairing) with TBG, as well as stacked and twisted TMDs.}
\label{tab1}
\begin{center}
\begin{tabularx}{\linewidth}{|X|X|X|X|}
\hline
{\bf Properties} & {\bf Twisted bilayer graphene} & {\bf Stacked and twisted TMDs} & {\bf Twisted Nodal Superconductors}\\
\hline
{\bf 
Monolayer electronic structure
}
& Semimetal; \;\;\;\;\;\; isotropic Dirac points at BZ$^{\rm a}$ corners &Semiconductor; \;\;\;\;\;\; Parabolic spin-valley locked hole bands at BZ$^{\rm a}$ corners & Superconductor; \;\;\;\;\;\; anisotropic BdG Dirac points on high symmetry lines \\[1.5cm]
{\bf
Effects of small twists 
}
& Isolated non-Wannierizable Flat bands near $\theta\approx \theta_{MA}$& Isolated Wannierizable Flat bands at low $\theta$ & No isolated minibands; Higher-order   BdG nodes at $\theta\approx \theta_{MA}$ \\[1.5cm]
{\bf
Effects of large twists 
}
& Structural quasicrystal near $\theta=30^{\circ}$ \cite{ahn2018dirac}
& Structural quasicrystal near $\theta=30^{\circ}$ \cite{tsang2024polar};  
A($\uparrow)$B$(\downarrow$) stacking at $60^{\circ}$~\cite{li2021quantum}&Spontaneous Time-reversal Symmetry breaking near $\theta_{TRSB}$ ($45^\circ$ for $d$-wave) \\[1.5cm]
{\bf 
Degrees of freedom (how to couple$^{\rm b}$)
}
&charge ($\mu$),\newline spin ($B_\parallel$),\newline valley ($B_\perp$),\newline layer($D$), \newline sublattice (hBN)
& charge ($\mu$),\newline valley/spin ($B_\perp$),\newline layer ($D$)  
&valley $I_\parallel$,\newline layer ($D$),\newline spin ($B_\parallel$),\newline $\arg \Delta $ ($I_\perp$, $B$),\newline  $|\Delta|$  ($T$, $B_\perp$) \\[2cm]
{\bf
Requirements for topological states
}
& alignment with hBN + valley polarization& Small twists of homobilayers;
Heterobilayers with $D$ 
& Application of $I_\perp$ at any $\theta$;  spontaneous near $\theta_{TRSB}$\\
\hline
\end{tabularx}
\end{center}
\begin{tabnote}
$^{\rm a}$ Brillouin zone --BZ; $^{\rm b}$ $D$ interlayer displacement field; $\mu$ - chemical potential tuned by voltage; $B_{\parallel(\perp)}$ - magnetic field; $I_{\parallel(\perp)}$ --- current, and $\Delta$ is the complex superconducting order parameter.
\end{tabnote}
\end{table}

Even less explored are the effects of strong interactions between BdG quasiparticles in the superconducting state. For nodal superconductors, competing interactions \cite{sachdev2002quantum} and phase fluctuations have been suggested to induce new phases \cite{herbut2002,franztes2001}; however, these effects have not been observed to date. Similarly, in monolayer graphene the interaction between Dirac electrons could potentially drive phase transitions, but fall short in strength \cite{ulybyshev2013}. For topological superconductors, interactions may lead to fractional or parafermionic excitations~\cite{volovik2000monopoles,Vaezi-PhysRevB.87.035132-2013,Mong-PhysRevX.4.011036-2014,Vaezi-PhysRevX.4.031009-2014,Sagi-Yuval-PhysRevB.96.235144-2017}, or ordered states of Majorana fermions\cite{rahmani2019interacting}. The realization of strongly interacting states of BdG quasiparticles thus serves as the next major frontier in superconductivity.



\begin{figure}[b!]
\begin{minipage}{0.45\textwidth}
    \includegraphics[width=\textwidth]{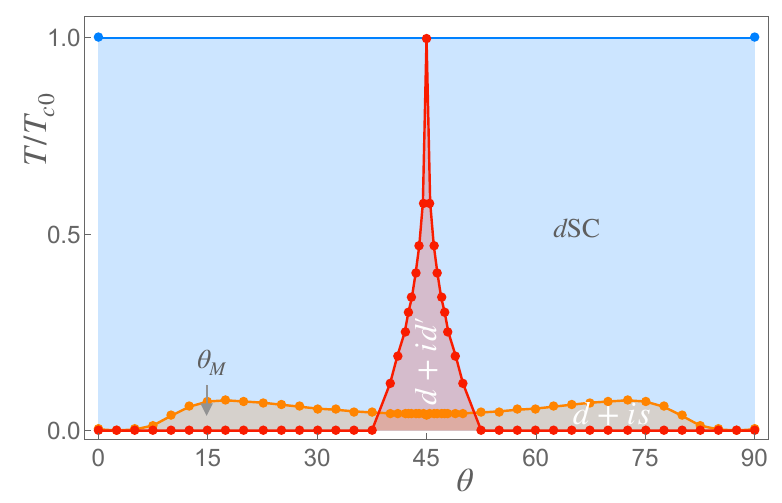}
\end{minipage}
\hfill
\begin{minipage}{0.5\textwidth}
    \includegraphics[width=0.5\textwidth]{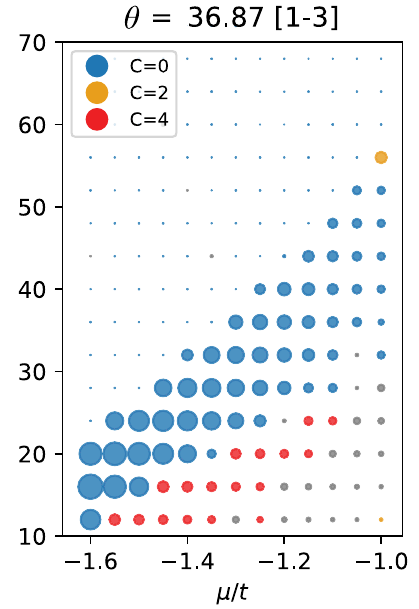}
\end{minipage}
\hfill
\begin{minipage}{0.45\textwidth}
    \includegraphics[width=\textwidth]{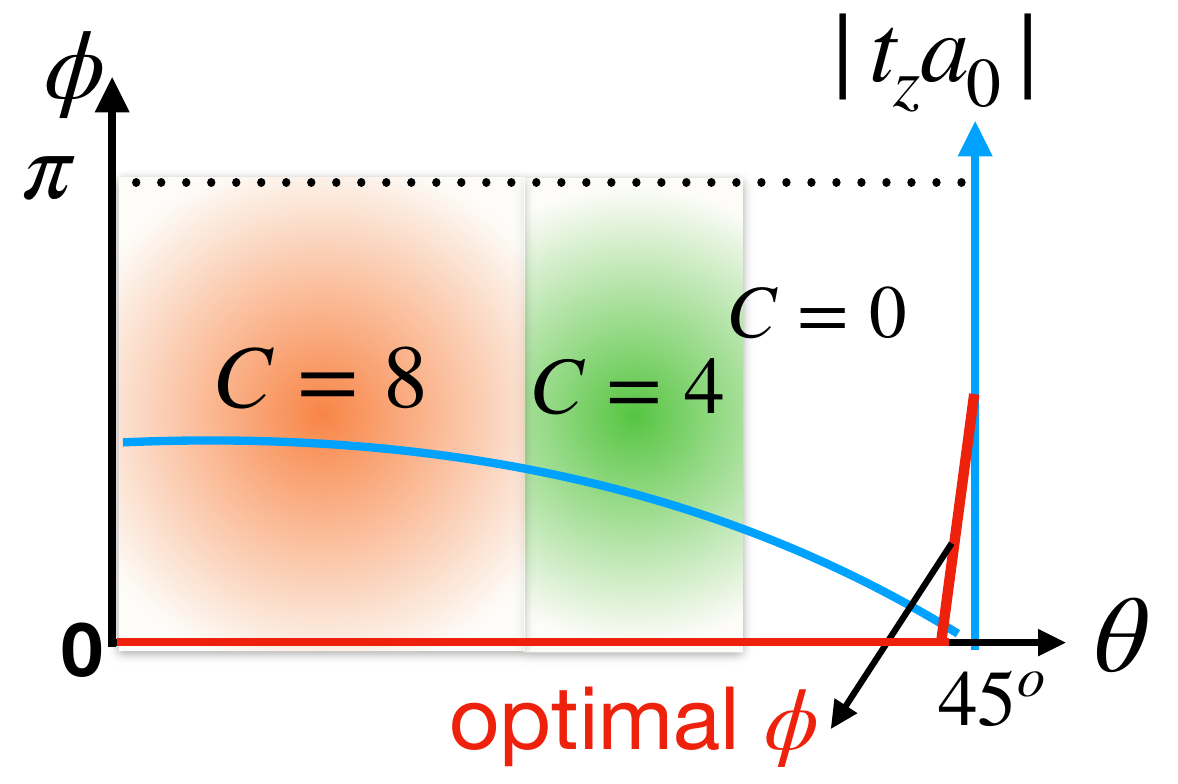}
\end{minipage}
    \begin{minipage}{0.45\textwidth}
    \includegraphics[width=\textwidth]{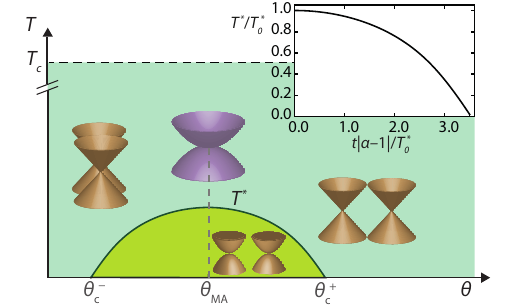}
\end{minipage}
\caption{{\bf Phase diagrams from large to small twist angles proposed for TNS.} (Top Left) Phase diagram in temperature (in units of $T_{c,0}$ of the monolayer) and twist angle, realizing a $d$-wave state (light blue region), a $d+is$ (orange region), and $d+id$ (red region) states~\cite{Tummuru-Franz-2022}. 
At smaller twist angles, near the magic-angle (see Sec.~\ref{sec:WF}) a topologically trivial time reversal breaking transition occurs, which is superceded by a topological $d+id$ state near twists of $45^\circ$.  This figure is reprinted with permission from \href{https://doi.org/10.1103/PhysRevB.105.064501}{Tummuru et. al, Phys. Rev. B 105, 064501 (2022).}
(Top right)  Phase diagram for a fixed twist angle  as a function of interlayer tunneling (vertical scale in unites of meV) versus twist angle $\theta$, the Chern number is marked by color and the size of the dotes denotes the minimum quasiparticle gap~\cite{can2021high}.
This figure is reprinted with permission from \href{https://www.nature.com/articles/s41567-020-01142-7}{Can et. al, Nature Physics 17, 519 (2021).}
(Bottom left) Schematic depiction of the topological phase diagram of a twisted $t-J$ model that describes the evolution of the Chern number expected for tBSCCO when including the full angular dependence of the interlayer tunneling in Eq.~\ref{eqn:interlayertunneling}~\cite{Song-2022}. This figure is reprinted with permission from \href{https://doi.org/10.1103/PhysRevB.105.L201102}{Song et. al, Phys. Rev. B 105, L201102 (2022).} (Bottom Right) Magic-angle induced instability in the BdG quasiparticle bands for circular Fermi surface due to the Dirac cones merging to form a quadratic band touching~\cite{VolkovPRB-2023}. (Inset) For a non-circular Fermi surface the time reversal breaking instability is gradually suppressed. To be contrasted with lattice model calculations  Ref.~\cite{Tummuru-Franz-PRB-2022}  shown in top left. Figure from Ref.~\cite{VolkovPRB-2023}.
}
\label{figpd}
\end{figure}

Twisted nodal superconductors appear to address all of these major challenges. We summarize the currently predicted emergent phenomena in Fig. \ref{figpd}, where the phase diagrams from a number of works on TNS are presented that range from large to small twist angles. Broadly speaking, stacking two nodal superconductors with a twist leads to two physical effects: hybridization of the order parameters (Sec.~\ref{sec:OP}) and of the quasiparticles (Sec.~\ref{sec:WF}). The first one can result in a spontaneous formation of a time-reversal symmetry breaking superconductor close to specific, but always large (e.g. $45^\circ$) twist angles \cite{manfred1992paramagnetic,kuboki1996,sigrist1998,can2021high} (Fig. \ref{figpd}, top and bottom left), and is insensitive to the thickness of individual superconductors (i.e. they need not be monolayers). This is in stark contrast to TBG and TMDs (Table \ref{tab1}), where at large twist angles and for interfaces between bulk, thick, flakes no emergent electronic physics is expected \cite{andrei2020graphene,mak2022semiconductor} (although structural  quasicrystals may form \cite{ahn2018dirac,tsang2024polar}). Quasiparticle hybridization, on the other hand, leads to similar effects as in TBG, quenching the velocity of the Dirac quasiparticles and enhancing the quasiparticle interactions that lead to phase transitions~\cite{VolkovPRL-2023,VolkovPRB-2023} (Fig. \ref{figpd}, top left and bottom right). The low-energy quasiparticles of TNS are highly tunable with current, magnetic, and electric fields (see Table \ref{tab1}), and may be manipulated with weak perturbations (relative to the maximum order parameter magnitude). This limits the pair-breaking effects when applying magnetic field and currents. Of particular mention is that time-reversal symmetry breaking (either spontaneous or induced by driving a non-dissipative current through the system) immediately leads to the realization of topological states (Fig. \ref{figpd}, top right and bottom left), not requiring any additional interaction effects, such as valley polarization in TBG~\cite{sharpe2019emergent,pixley2019ferromagnetism,serlin2020intrinsic} or TMDs \cite{mak2022semiconductor}.

The initial impetus for the TNS field is partly attributed to the realization of monolayers of the high-$T_c$ cuprate Bi$_2$Sr$_2$CaCu$_2$O$_{8+x}$ (BSCCO) \cite{yu2019}, which remains the main tool for experimental progress \cite{zhao2023,xue_2021,xue_2023_OP,martini2023twisted,lee2023encapsulating,confalone2025challenges}. We will thus dedicate special attention to twisted $d$-wave spin singlet superconductors, including both theoretical calculations and  its experimental realization in Josephson junctions of twisted \bscco (tBSCCO). On the other hand, one of the goals of the present review is to motivate the development of other material platforms for TNS.

The rest of the paper is organized as follows. In Sec.~\ref{sec:OP} we review the effects of hybridization between the order parameters in TNS. We focus on the theoretical predictions of phase-sensitive order parameter probing and broken time reversal symmetry and the evolution of experimental efforts on c-axis twist Josephson junction in \bscco over the last 26 years. In Sec.~\ref{sec:WF} we review the physics of quasiparticles in TNS, including magic angle effects and emergent correlated and topological phases, as well as proposals for experiments (Sec. \ref{subsec:topexp}) that can probe topological superconductivity in TNS. We summarize and discuss future directions in Sec.~\ref{sec:conclusions}.

\section{Order parameter hybridization}
\label{sec:OP}
\begin{figure}[h]
\includegraphics[width=0.9\textwidth]{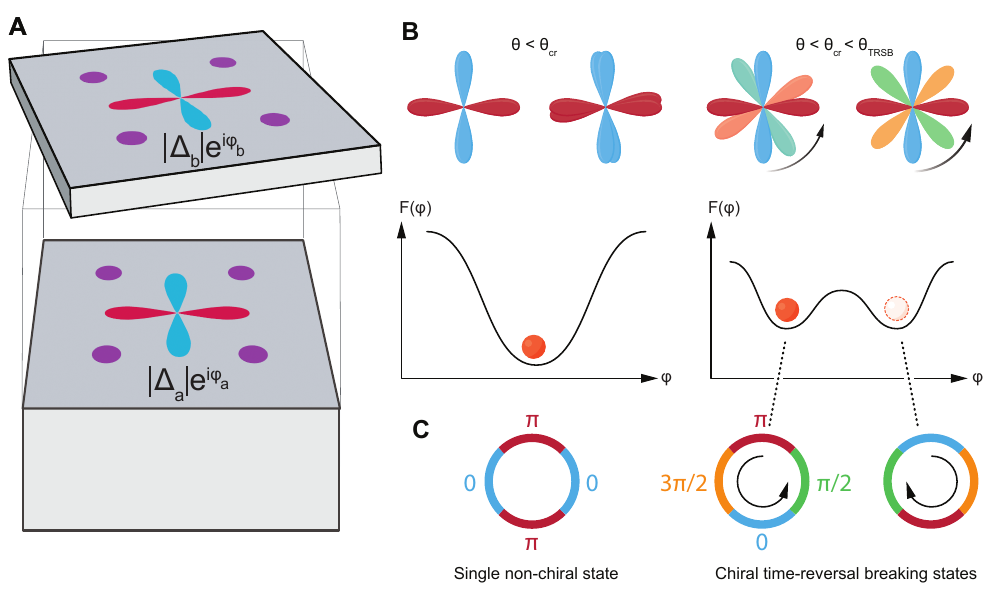}
\caption{{\bf Order parameter hybridization in twisted nodal superconductors.} ({\bf A}) Twisted interface between two bulk nodal superconductors. The relative phase of the order parameters is determined by the coupling between order parameters across the interface. ({\bf B}) schematics of the two order parameters and interfacial free energy as a function of phase difference $\varphi=\varphi_a-\varphi_b$. At low twist angles, the optimal state has $\varphi = 0$. In the vicinity of $\theta_{TRSB}$ ($45^\circ$ for $d$-wave), in contrast, there are two inequivalent free energy minima with nonzero phase difference. Black arrows on the top right indicate the preferred chirality of the state. ({\bf C}) Shows the evolution of the SC order parameters' phases (from panel {\bf B}) as a function of in-plane azimuthal angle. In the time-reversal breaking case, the phase increases by going clockwise or anticlockwise for the two possible states, while for $\theta<\theta_{cr}$ the phase oscillates between $0$ and $\pi$ without a preferred direction.
}
\label{fig2}
\end{figure}

In this section, we focus on the physics of the order parameter, see Fig.~\ref{fig2}. Its presence is the main feature that distinguishes TNSs from other twistronic platforms (see Table~\ref{tab1} for a comparison of TNS with TBG and twisted and stacked TMDs). Moreover, a self-contained description of the order parameter physics is possible within phenomenological Landau theory \cite{SigristUeda,can2021high,volkov_diode}. 
Importantly, we will focus on order parameters where nodes are symmetry-enforced (e.g., $d,p$-wave pairing), rather than accidental (like extended $s$-wave pairing).

Nodal order parameters (such as the $d_{x^2-y^2}$ one depicted in Figs.~\ref{fig1} and \ref{fig2}) are typically not invariant under rotation. With the lobes of the order parameter fixed to the crystalline axes, the order parameter is bound to rotate together with the flake, allowing mechanical control over superconductivity. 
This observation has motivated proposals for phase-sensitive tests of the order parameter symmetry \cite{geshkenbein1987,harligen1995,Tsuei-RevModPhys.72.969-2000}. As we show below, more recent developments, partially inspired by those ideas, open the potential not only to probe, but to engineer superconducting orders in TNS.

The basic setup we will consider in this section is shown in Fig. \ref{fig2} and consists of two flakes of a nodal superconductor stacked on top of one another with a twist. Remarkably, for the results discussed in this section the flakes need not be monolayers; in fact they can  as well be bulk superconductors. Each separate flake hosts a uniform superconducting order parameter $\Delta_{a,b} = |\Delta_{a,b}| e^{i\varphi_{a,b}}$, characterized by an amplitude $|\Delta_{a,b}|$ and a phase $\varphi_{a,b}$. In the absence of coupling between flakes, the amplitudes $|\Delta_a| =|\Delta_b|$ are fixed by the interactions within each flake, while $\varphi_{a,b}$ are not. As the global $U(1)$ phase is not observable, we can take $\varphi_a+\varphi_b=0$ without loss of generality, while leaving $\varphi_a-\varphi_b=\varphi$ undetermined. The most important role of the coupling between the TNS across the interface is thus to fix the {\it relative phase} $\varphi$ of the order parameters in the two flakes. One may thus expect the interfacial contribution to the free energy to be a function of $\varphi$, that can have one or more minima (see Fig. \ref{fig2} B).

We now set specific expectations using Landau theory. While using this approach for illustrative properties, we remark that the results of this analysis qualitatively agree with more microscopic calculations \cite{can2021high,volkov_2025,Song-2022,senechal_2022}.
Consider the leading terms of the Landau expansion that explicitly depend on the phase difference $\varphi$ between the order parameters:
\begin{equation}
    F_{inter} = \alpha(\theta)[\Delta_a^*\Delta_b+c.c.] + \beta(\theta) [(\Delta_a^*)^2(\Delta_b)^2+c.c.]  = -A(\theta,T) \cos \varphi+ B(\theta, T) \cos 2\varphi,
    \label{eq:F}
\end{equation}
where coefficients $A,B$ depend on temperature and the twist angle. The interface free energy is related to a directly measurable quantity --- the supercurrent flowing through the interface by \cite{golubov2004}:
\begin{equation}
    I(\varphi) = \frac{2e}{\hbar} \frac{d F_{inter}(T,\varphi)}{d\varphi} = A(\theta,T) \sin \varphi-2 B(\theta, T) \sin 2\varphi.
    \label{eq:cur}
\end{equation}
In particular, the critical current that the interface can sustain is determined as $I_c(\theta) = \max_\varphi I(\varphi)$.

The twist angle dependence in Eq.~\eqref{eq:F} is what distinguishes between TNS and ordinary superconductors. For usual superconductors weakly coupled across an interface, $A>0,\;A\gg B$ and $\varphi=0$ is the ground state (Fig. \ref{fig2}). Moreover, for perfectly isotropic $s$-wave pairing and a circular Fermi surface, one expects no angular dependence at all (although for a more realistic one, some angular dependence is expected, see below).

In contrast, for TNS, the dependence of Eq.~\eqref{eq:F} on angle is unavoidable. Let us take $d_{x^2-y^2}$ pairing on a square lattice (point group $D_{4h}$) as an example. Its main characteristic is the sign change under $90^\circ$ rotation $C_{4z}$: $C_{4z} \Delta_{a,b} = -\Delta_{a,b}$. In the geometry of Fig. \ref{fig2}, $C_{4z}$ remains a symmetry at all twist angles. However, at $\theta=45^\circ$, the system's symmetry is enlarged by an operation combining $45^\circ$ rotation and inversion : $S_8$ \cite{volkov_diode}. Since the order parameters are already rotated by $45^\circ$, this operation acts on order parameters as $\Delta_a \to \Delta_b,\Delta_b \to -\Delta_a$. The presence of such symmetry explicitly implies $\alpha,A=0$, in stark contrast with the ordinary case discussed above. Pictorially, rotating the $d_{x^2-y^2}$ order parameter in one layer by $45^\circ$ transforms it into a $d_{xy}$ one, which is orthogonal to  $d_{x^2-y^2}$. 
This argument assumes a weak interfacial coupling that does not affect the pairing states within each flake; for stronger coupling the same conclusions can be reached by analyzing the pairing symmetries for the $D_4$ point group of twisted bilayer \cite{senechal_2022}. Qualitatively, one may expect a strong suppression of the critical current $I_c$ near $45^\circ$, as the second-harmonic term is typically weaker, arising from higher-order processes~\cite{Tummuru-Franz-PRB-2022,volkov_2025} or inhomogeneity \cite{yuan2023,yuan2023-2,yuan2024}.

The resulting ground state near $\theta=45^{\circ}$ is then determined by the second term in \eqref{eq:F}. Microscopic calculations \cite{Tummuru-Franz-PRB-2022,yuan2023,volkov_2025,senechal_2022}, suggest $B>0$, leading to the existence of two minima at $\varphi=\pm\pi/2$ (Fig. \ref{fig2} B,C). A finite phase difference across the junction is generated and breaks time reversal symmetry spontaneously. Small deviations from $45^\circ$ twist will not immediately destroy the time-reversal broken state \cite{can2021high,volkov_2025}, apart from the immediate vicinity of $T_c$, where in Eq.~\eqref{eq:F}, $A(\theta,T)\propto (T-T_c)$ and $B(\theta,T)\propto (T-T_c)^2$, if allowed. The resulting phase diagram in Fig.~\ref{figpd} (Left) then forms a wedge in twist-temperature space, where the TRSB state extends to $T_c$ only at $45^\circ$ (and therefore define $\theta_{TRSB}=45^\circ$ for the $d$-wave case). Qualitatively similar phase diagram has been found for grain boundary junctions of $d$-wave superconductors with changing crystal orientation \cite{yip1993,yip1995,kuboki1996,sigrist1998}. The difference between the two geometries becomes critical in the limit of two twisted monolayers: while the in-plane junction would form along a 1D line, the c-axis junction will cover the whole 2D material, affecting the bulk quasiparticle properties in a profound way, as discussed in Sec.~\ref{sec:WF}.

The discussion above is straightforward to generalize to any other symmetry-protected nodal superconductors. For a given lattice and pairing symmetry, a twist angle $\theta_{TRSB}$ can be found, where the emergent $S_n$ symmetry (e.g. $S_{6}$ for 30$^\circ$ twisted $D_{3h}$ materials \cite{zhou2023non} or $S_4$ --- for 90$^\circ$ twisted $D_{2h}$ ones \cite{Tummuru-Franz-2021}) excludes the first-harmonic Josephson tunneling, allowing for a TRSB state. This effect does not depend on whether the pairing is triplet or singlet - the only consideration is that of symmetry.

\subsection{Twist junctions as a probe of pairing symmetry}
Having set the general expectations, we now review how much of those expectations bore out under theoretical and experimental scrutiny. We will begin with discussing how the pairing symmetry can be established via the twist angle dependence of the critical current.

The motivation appeared quite naturally in the field of cuprates, where the c-axis twist between layers appeared to be a defect, naturally occurring in BSCCO \cite{qli1998}. This material is characterized by the highest degree of two-dimensional anisotropy among the cuprates, allowing for the isolation of a single monolayer~\cite{yu2019} or few-layer thick films \cite{Zhao-Kim-2019}. Structurally, it preserves the $C_4$ symmetry \cite{klemm2005phase}, unlike YBa$_2$Cu$_3$O$_7$~\cite{Tsuei-RevModPhys.72.969-2000}, ensuring that the gap nodes should occur along the Brillouin zone diagonals (although signatures of electronic nematicity and charge density waves that can break $C_4$ have been reported \cite{keimer2015,proust2019}).

\begin{figure}[h]
\includegraphics[width= \linewidth]{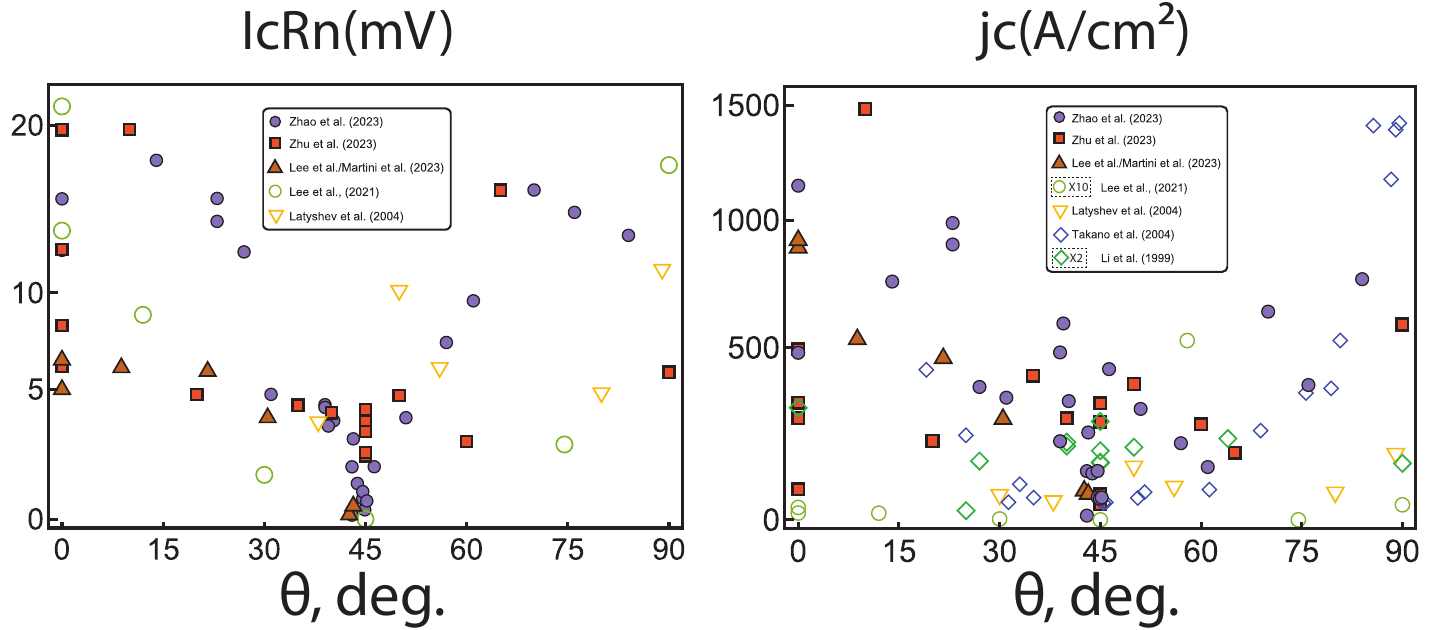}
\caption{Critical current as a function of the twist angle. Left: product $I_cR_N(\theta)$ is shown, right: critical current density is shown. All data is from optimally doped \bscco twist junctions at the temperatures given below. Empty symbols are for devices that required any annealing, full symbols are for devices prepared using cold tear-and-stack technique that required no annealing. Data taken from: Zhu et al. (2023),$T=1.6$ K \cite{xue_2023_OP}; Zhao et al. 2023, $T=30$ K\cite{zhao2023}; Lee et al. and Martini et al. (2023), $T=5$ K \cite{lee2023encapsulating} \cite{martini2023twisted}; Lee et al. (2021), $T=4.8$ K \cite{lee2021twisted}; Takano et al. (2004), $T=5$ K\cite{takano_2004}; Latyshev et al. (2004), $T=4.2$ K \cite{latyshev2004}; Li et al. (1999), $T=0.9 T_c$ \cite{qli1999}. In two cases a multiplicative factor is used to help to see all of the data on a common scale.}
\label{fig3}
\end{figure}

The initial results \footnote{Some data on such systems has been obtained earlier \cite{tomita1992,wang1994}.} on the artificial twist junctions \cite{qli1999} (Fig. \ref{fig3} right, green empty diamonds) have demonstrated a complete absence of twist angle dependence. We would like to stress
that $I_c(\theta = 45^\circ)$ is not required (or expected) to vanish exactly even for a pure $d$-wave superconductor junction due to the presence of a second-order coupling~\cite{volkov_2025}, in Eq. \eqref{eq:F}. Nonetheless, since the latter term is expected to be small, a complete absence of angular dependence is not expected for the $d$-wave case. Subsequent experiments on junctions made from ``whiskers" of BSCCO (that grow along the in-plane $a$ axis) \cite{takano_2002,takano_2004} have shown a strong variation of $j_c$ with a minimum near $45^\circ$ (blue empty diamonds), which was later contrasted by the results on naturally occurring cross-whisker junctions \cite{latyshev2004} (empty yellow upside-down triangles) interpreted in favor of $s$-wave pairing. One common characteristic of all the experiments in this group is the necessity to anneal the samples (empty symbols are used for annealed devices): the ones in \cite{qli1999} were held just below the melting temperature for 30 hours, while both whisker experiments \cite{takano_2002,latyshev2004} required annealing in an oxygen atmosphere at 850 $^\circ$C for 30 minutes and 845 $^\circ$C for 20 minutes, respectively.

Over time, evidence from multiple probes accumulated for $d$-wave, nodal superconductivity in \bscco and cuprates in general: (see \cite{fujita2011qpi,hashimoto2014arpes} and references therein) including penetration depth \cite{sonier2000,prozorov2006}, heat transport \cite{shakeripour2009heat}, 
corner and grain boundary junctions \cite{harligen1995,Tsuei-RevModPhys.72.969-2000}, and Andreev bound state spectroscopy \cite{kashiwaya2000, lofwander2001}. One exception was the observation of a finite, first-order, Josephson effect between Pb and BSCCO \cite{kleiner1999pb}, which should be symmetry-forbidden for a pure $d$-wave superconductor. However, the value of the critical current observed was small, corresponding to a one-thousands admixture of $s$-pairing at most, also inconsistent with predominantly $s$-wave pairing.



The c-axis twist junction experiments have seen a resurgence lately due to rapid development of tear-and-stack techniques \cite{geim2013van}.
The resulting junctions were made of thinner flakes than before, and required (optimally- \cite{lee2021twisted} and under-doped \cite{xue_2021}) less annealing at lower temperature in oxygen (350$^\circ$C for 30 min  and  530$^\circ$C for 10 min, respectively).
We note, however, that for these devices the critical temperature of the junction has been noticeably lower than that of the flakes separately (which was comparable to the bulk $T_c$ at the corresponding doping), in contrast to \cite{qli1999}, while individual monolayers of \bscco show the same $T_c$ as the bulk \cite{yu2019}. 


The next advance has been the development of cryogenic assembly of twisted BSCCO flakes \cite{zhao2023}, which allowed exfoliation and assembly while keeping the sample at  -100$^\circ$C during the process. The resulting junctions required no annealing to 
superconduct and have the junction critical temperature close to that of the bulk. The $I_c(\theta)$ data (purple filled circles \cite{zhao2023} and brown filled triangles \cite{martini2023twisted,lee2023encapsulating}, Fig. \ref{fig3}) show a pronounced variation with twist angle and roughly follows $I_c \propto \cos 2\theta$ \cite{zhao2023,martini2023twisted,lee2023encapsulating} consistent with Eq.~\eqref{eq:cur}. Data on samples obtained with the same method by other groups has also been reported (red filled squares, Fig. \ref{fig3}) \cite{xue_2023_OP} and interpreted in terms of an anisotropic $s$-wave pairing.

While different experiments appear mostly consistent regarding the order of magnitude, all show some sample-to-sample variations.
Specifically, $j_c$ appears to have more variation than $I_cR_N$. While both quantities are expected to be size-independent and therefore probe an average bulk property, they contain different uncertainties. $I_cR_N$ for nodal superconductors depends on the degree of momentum conservation during tunneling (see Sec.\ref{subsec:OP:theor}) and thus may fluctuate when the interface roughness or disorder varies. On the other hand, $j_c$ estimates use an optically determined area. The presence of strongly insulating regions of the flake, where local resistance is large and $j_c$ is extremely low or zero, would lead to variations. 
The larger variations of $j_c$ suggest that local high-resistance regions may be important.

From the above discussion, it becomes clear that the \bscco interfaces are quite sensitive and may be affected by the sample preparation method. One potential reason is the incommensurate structural supermodulation in the insulating BiO layers (see \cite{poccia2011} and references therein). However, 
transmission electron microscopy \cite{qli1998,lee2021twisted,xue_2021,zhao2023} has consistently 
demonstrated good quality and high degree of interface ordering.



\subsubsection{The vicinity of $\theta=45^\circ$ and the current-phase relation}
We would like to emphasize that the qualitative difference between the $d$- and $s$- wave case is not the absence of critical current at $45^\circ$, but the vanishing of the {\it first harmonic} of the critical current, Eq. \ref{eq:cur}. The implication is that $I_J(\varphi) \propto \sin 2 \varphi$ at $\theta=45^\circ$ is the defining feature of the $d$-wave case. This can be probed experimentally by analyzing the in-plane magnetic field dependence of the critical current, $I_c(B_\parallel)$. In short (smaller than the Josephson length \cite{tinkham2004introduction,volkov_2025}) junctions, the expected result is $I_c(B_\parallel) = I_{c0} \frac{\sin \frac{2\pi \Phi}{\Phi_0}}{\frac{2\pi \Phi}{\Phi_0}}$ and $I_c(B_\parallel) = I_{c0} \frac{\sin \frac{\pi \Phi}{\Phi_0}}{\frac{\pi \Phi}{\Phi_0}}$ for first- and second- harmonic current-phase relation, respectively, where $\Phi = B_\parallel A_\parallel$ is the magnetic field flux through the junction and $\Phi_0 = \frac{\pi \hbar c}{e}$ the flux quantum. Due to irregular shapes of the flakes and Meissner currents, determining the effective junction area is not trivial \cite{zhao2023,volkov_2025}. In \cite{zhao2023}, two junctions with very similar thicknesses and areas, at and away from $45^\circ$, did show a difference of $2$ in the period of $I_c(B_\parallel)$ oscillations, with the device closer to $45^\circ$ showing the doubled period. An independent way to assess the presence of the second harmonic are fractional Shapiro current steps, appearing in junctions driven by an AC field at frequency $\omega$ at voltages $V_n^{(2)} = \frac{n \hbar \omega}{4e}$ (opposed to $V_n= \frac{n \hbar \omega}{2e}$ in the first harmonic case). Both experiments have demonstrated the presence of the second harmonic in the current-phase relationship in Ref.~\cite{zhao2023}. In contrast, previous experiments have shown only the conventional Shapiro steps \cite{takano_2004,lee2021twisted}.



Before we move on to discussing the theoretical developments largely motivated by these experiments, we would like to highlight that the van-der-Waals route to creating twisted interfaces between superconductors has been applied to wider classes of materials: other cuprates, such as Bi$_2$Sr$_{2-x}$La$_x$CuO$_{6+y}$  \cite{xue_2023_bi2201,zhang2023josephson} and  different van der Waals superconductors, such as NbSe$_2$\cite{yabuki2016supercurrent} and NbS$_2$\cite{zhao2022josephson} that will be discussed in more detail in Sec. \ref{sec:conclusions}. In the former case, \cite{xue_2023_bi2201}, the temperature dependence of the critical current has been suggested as evidence for dominant first harmonic near $45^\circ$ twist, while the observed Fraunhofer patterns and Fiske (induced by standing electromagnetic waves) steps appeared consistent with both.


\subsubsection{Theoretical description}
\label{subsec:OP:theor}
The Landau theory allows one to make qualitative predictions only in the vicinity of certain twist angles (e.g. 45$^\circ$ for $d$-wave) and sufficiently close to $T_c$. Extending the theory to describe $I_c(\theta, T)$ for all $\theta, T$ requires microscopic models. Moreover, one has to go beyond the standard Ambegaokar-Baratoff theory \cite{ambegaokarbaratoff} that assumes incoherent (i.e. with equal amplitude for tunneling between any two momenta of the two layers) interfacial tunneling, as it leads to zero critical current for TNS. Early theory \cite{tanaka1997} used a continuum anisotropic 3D model for the bulk of the flakes, however for highly two-dimensional materials a better picture is to view the whole material as a stack of weakly Josephson coupled layers \cite{klemm2005phase}. Indeed, for \bscco a strong motivation comes from the observation of the Josephson effect in bulk single crystals (see \cite{kleiner1997intrinsic} and references therein). For the case of momentum-conserving tunneling, the critical current gets rapidly suppressed at nonzero $\theta$ due to a reduced overlap of the Fermi surfaces \cite{klemm2005phase,Tummuru-Franz-PRB-2022,volkov_2025}. This principle applies even to $s$-wave superconductors. For example, twisted NbSe$_2$ flakes (where the bulk is believed to be $s$-wave, see however the discussion  in Sec.~\ref{sec:conclusions}) show a pronounced $I_c(\theta)$ minimum at $30^\circ$ \cite{yabuki2016supercurrent,farrar2021superconducting}, that may be attributed to this effect since there the Fermi surface mismatch is maximal. Returning to BSCCO, the rather smooth $I_c \propto \cos 2 \theta$ dependence observed in \cite{zhao2023} suggests a degree of momentum relaxation \cite{volkov_2025}, pointing to a role of inhomogeneity. In the earlier studies, a possibility of mixed order parameters in individual flakes \cite{klemm2005phase} was considered. For a mixing induced by broken rotational symmetry, e.g. $d_{x^2-y^2}+s$, the zeroes of the lowest-order Josephson current can shift away from $45^\circ$, whereas for time-reversal breaking mixtures (e.g., $d_{x^2-y^2}+is$ or $d_{x^2-y^2}+id_{xy}$) a complete avoidance of zeroes is possible \cite{klemm2005phase}.

Recent work focused on exploring the more detailed features of the Josephson effect, such as nonmonotonic temperature dependence \cite{Tummuru-Franz-PRB-2022,volkov_2025}, observed in \cite{zhao2023}, and second harmonic contributions to the Josephson current. For the latter, two potential mechanisms were identified: Cooper pair cotunneling arising from higher-order corrections from tunneling \cite{Tummuru-Franz-PRB-2022,volkov_2025} and an inhomogeneity-induced mechanism \cite{yuan2023,yuan2023-2,yuan2024}, where either twist angle disorder or electronic nematicity can lead to a sign-changing spatially inhomogeneous profile of the Josephson critical current density near 45$^\circ$. Another interesting aspect of relevance to cuprates is the potential effect of charge density waves \cite{keimer2015,proust2019} on Josephson tunneling \cite{gabovich2023coexistence} under twist. Effects of correlations within layers \cite{Song-2022,senechal_2022} and momentum-dependent tunneling \cite{volkov_2025} have been shown to lead to a similar qualitative picture. Nonetheless, quantitative agreement with experimental results \cite{zhao2023,xue_2023_review} has not yet been achieved.

The resurgence of activity on twisted cuprate interfaces has motivated applying  twist junctions to probe the pairing symmetry in other systems and inspired new approaches, e.g. using momentum-resolved tunneling to probe unconventional van der Waals superconductors  \cite{xiaoberg_2023}  or using edge effects to extract the full orientation dependence of the order parameter 
\cite{yuan2024phase}. These ideas are especially relevant for 2D materials, where many conventional thermodynamic probes of superconductivity are not available, while c-axis twist junction geometry is natural.

\subsection{Time reversal symmetry breaking}
\label{subsec:trsb}

A major qualitative prediction of the phenomenological theory, Eq.~\eqref{eq:F} is the presence of a time-reversal breaking phase near $\theta_{TRSB}$ when $B>0$. The latter, however, is not required by symmetry and has to be calculated microscopically. In this regard, different theoretical treatments for the $d$-wave case uniformly predict that $B>0$ \cite{can2021high,Tummuru-Franz-PRB-2022,senechal_2022,Song-2022,yuan2023} (one possible exception is for strong momentum relaxation at the interface \cite{volkov_2025}). This prediction has been found to extend to other systems such as $p$- and $f$-wave superconductors \cite{Tummuru-Franz-2021,zhou2023non}. Moreover, it has been demonstrated that the order parameter fluctuations can drive transitions into composite ordered phases above the bulk superconducting $T_c$ near $\theta_{TRSB}$ \cite{liu2023charge}.

Experimental evidence for time-reversal breaking superconductivity in t\bscco  flakes has been provided by the Josephson diode effect \cite{nadeem2023superconducting}, where the critical current is non-reciprocal. This effect is forbidden in systems with time-reversal symmetry, but has been found to occur in close-to-45$^\circ$ t\bscco flakes in the absence of a magnetic field at low temperatures \cite{zhao2023}. Theoretically, the Josephson diode effect has been shown to be a sufficient, but not a necessary condition for time-reversal breaking \cite{volkov_diode}; moreover, the diode direction should be trainable by current for spontaneously broken symmetry (as in Fig. \ref{fig2}). In experiments \cite{zhao2023} it has been found that while current training affects the diode direction, it does not reverse the effect completely, and this is not altered with thermal cycling above $T_c$\cite{zhao2023}. 
This behavior is consistent with an additional time-reversal symmetry breaking~\cite{volkov_diode} unrelated to superconductivity. One candidate for the origin of time-reversal symmetry breaking is the pseudogap phase \cite{keimer2015,proust2019}; one way of testing this hypothesis is considering overdoped samples, where the pseudogap is suppressed. On the other hand, the time-reversal symmetry breaking may be occurring at the interface only, caused by an instability there. This raises an intriguing possibility that the diode effect can be a probe of the enigmatic correlated phase diagram of the cuprates.

The superconducting diode effect has by now been observed by several groups. Ref.~\cite{xue_2023_OP} reported a field-free diode effect at 45$^\circ$  tBSCCO interfaces. Theoretically the diode effect should vanish exactly at $45^\circ$ \cite{volkov_diode} due to a mirror symmetry (in $\varphi$) around each free energy minimum. Additionally, a magnetic field-induced diode effect has been reported in Ref.~\cite{ghosh2024high}. Finally, the superconducting diode effect in the absence of a magnetic field  has been recently reported in single flakes of \bscco~\cite{diode_1flake_2025}. However, care needs to be taken to avoid signals due to trapped vortices \cite{ghosh2024high}; e.g., in \cite{zhao2023} a cryostat without a magnet was used for diode effect measurements.

This calls for alternative probes of time-reversal symmetry breaking at twisted interfaces to be employed; the recent proposals suggest using polar Kerr effect \cite{can_pke}, probing spontaneous currents at edges~\cite{franzedge} or around impurities ~\cite{jxzhu_2024}, and non-linear response with THz light~\cite{kaplan2025quantum}. Optical probes of t\bscco are also actively being developed experimentally \cite{xiao2024optically}.

\begin{summary}[SUMMARY POINTS: Order Parameter Hybridization]
\begin{enumerate}
\item Destructive order parameter hybridization near $\theta_{TRSB}$ (e.g., 45$^\circ$ for $d$-wave) leads to dominant higher-harmonics in the current-phase relation and can induce spontaneous time-reversal symmetry breaking.
\item c-axis twist interfaces can serve as a phase-sensitive probe of the order parameter for both 2D (van der Waals) and bulk materials
\item The most studied platform at present are \bscco flakes; cryogenic preparation of samples in inert atmosphere is important for the interface quality. The role of correlations and non-superconducting electronic orders on the superconducting order parameter hybridization remains an open question. 
\end{enumerate}
\end{summary}

\section{Quasiparticle hybridization in twisted nodal superconductors}
\label{sec:WF}

\begin{figure}[h]
\includegraphics[width= \linewidth]{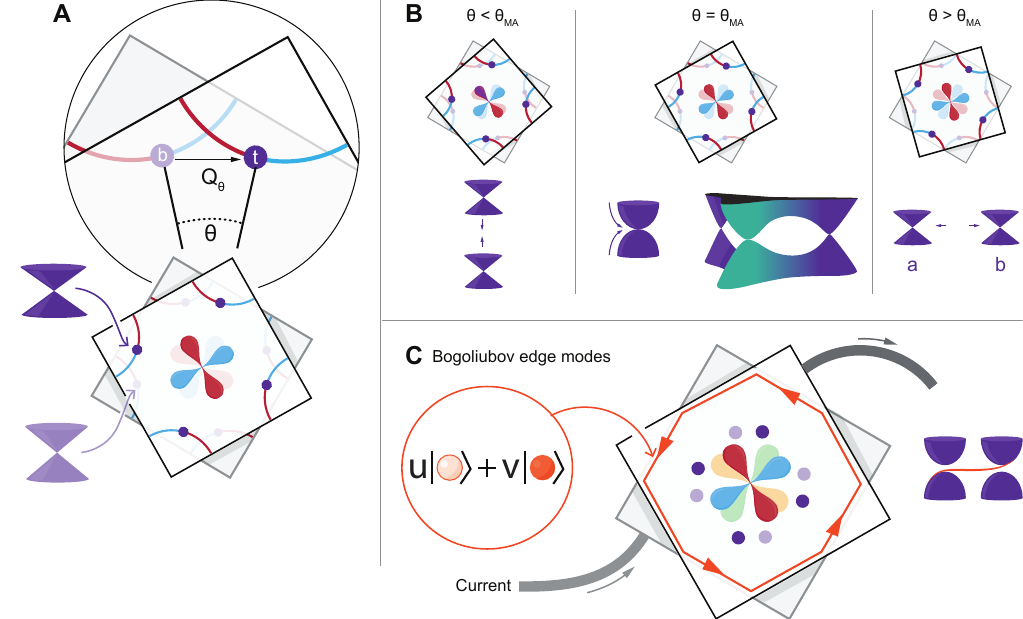}
\caption{{\bf Quasiparticle  hybridization in twisted nodal superconductors.} {\bf A} Anisotropic Dirac cones in the two layers separate in momentum space due to the twist $\theta$ (marked by purple disks) by $Q_{\theta}$. Interlayer hybridization of quasiparticles ({\bf B}) leads to a reconstruction that flattens the dispersion and brings the Dirac points together. At $\theta \approx \theta_{MA}$ (not to be confused with $\theta_{TRSB}$ in Sec.~\ref{sec:OP}), the Dirac dispersion softens completely, driving symmetry-breaking instabilities (see Fig.~\ref{figpd}, top left and bottom right). ({\bf C}) In the presence of an interlayer current (or spontaneous time reversal symmetry breaking discussed in Sec. \ref{sec:OP}), the Dirac points gap out, realizing a topological superconductor with gapless BdG edge modes (orange lines and band connecting the gapped bulk bands to the right) that are a linear combination of particles (lightly colored orange sphere) and holes (darkly colored orange sphere). We note that for an actual experiment, the non-overlapping parts of the device need to be removed, to avoid the presence of gapless Dirac quasiparticles, or cuts in the interior of the flake (hosting edge modes) created \cite{ji2024local}.
}
\label{fig4}
\end{figure}

In this section we will review the physics of the BdG quasiparticles in TNS (Fig. \ref{fig4}). We will set the stage by reviewing TNS bilayers at low twist angles ~\ref{subsec:lowen}, where hybridizing BdG Dirac quasiparticles will allow us to establish analogies to TBG. A distinctive feature of TNS is the emergence of topological states under broken time-reversal symmetry \cite{can2021high,VolkovPRL-2023} that we discuss in \ref{sec:subsec:topology}. While time-reversal symmetry can be broken spontaneously, as described in the preceding section, breaking time-reversal externally, e.g. by applying a current (Fig. \ref{fig4}, C) is sufficient to realize topological superconductivity in TNS {\it at any twist angle}. In Sec.~\ref{subsec:correlations} we will discuss the effects of interactions between quasiparticles and the mutual influence of order parameter and the quasiparticles.  At present, the properties of TNS reviewed in this section remain largely a theoretical effort; we have collected  the proposed experimental signatures of teh topological states in Sec. \ref{subsec:topexp} and show how to tune TNS quasiparticles in analogy with TBG and TMDs in Table~\ref{tab1}.

\subsection{Low twist angles: the "magic angle"}
\label{subsec:lowen}

We begin by reviewing the low-energy description \cite{VolkovPRB-2023} of twisted bilayers of nodal superconductors in the mean-field approximation. This assumes long-range superconducting order, which occurs for a 2D system only for $T=0$ in the strict sense, but the emergent gapped phases may be expected to be robust against phase fluctuations at low temperatures. In general, the mean field BdG Hamiltonian of a single layer  can be written as $H_{BdG}^l({\bf K})=(\epsilon({\bf K})-\mu) \tau_3 +\hat{\Delta}({\bf K})$ is 
a $4 \times 4$ matrix acting in particle-hole (Gor'kov-Nambu) and spin space, denoted by Pauli matrices $\tau_i$ and $s_i$, respectively. We denote the layer by $l=t,b$, the single particle dispersion as $\epsilon({\bf K})$, while $\hat{\Delta}({\bf K})$ is the superconducting order parameter, which can be an arbitrary matrix off-diagonal in Gor'kov-Nambu space. In this section, we will not consider the self-consistent formation of the order parameter and assume its value (but not phase) to be fixed in each layer separately; this constitutes a good approximation when the interlayer tunneling ($t$, defined below) is much less than the maximal magnitude of $\hat{\Delta}({\bf K})$~\cite{VolkovPRB-2023}, for an extended discussion see Sec. \ref{subsec:correlations}.

In a nodal superconductor, the order parameter vanishes at certain points (${\bf K}_N$) on the Fermi surface (Fig. \ref{fig4} A). In their vicinity, a linear expansion of the $\epsilon({\bf K})$ and $\hat{\Delta}({\bf K})$ results in (note that there are several distinct nodes usually, e.g. 4 for d-wave, Fig. \ref{fig4} A):
\begin{equation}
\mathcal{H}_{BdG}^l({\bf K}_N+ {\bf k}) \approx {\bf v}_F \cdot {\bf k}  \tau_3+ {\bf v}_\Delta \cdot {\bf k}  \hat{\Delta}({K}_N),
    \label{eqn:hamnotwist}
\end{equation}
where ${\bf v}_F \perp {\bf v}_\Delta$ for symmetry-enforced nodes (such as in the $d$-wave case) \cite{VolkovPRB-2023}.
The twist results in the Dirac nodes of two layers being no longer at the same momentum, i. e. $K_N^{top} \neq K_N^{bot}$ (see the purple disks in Fig. \ref{fig4} A). 

Hybridization between the quasiparticles of the two layers arises from single-electron tunneling. The interplay between the twist, separating the Dirac point, and tunneling calls for an analogy with twisted bilayer graphene. A crucial difference, however is that Dirac points in superconductors do not generically occur at high-symmetry points of the Brillouin zone. As a result, there is a single dominant tunneling process with momentum exchange ${\bf Q}_\theta = \hat{z}\times{\bf K}_N\theta$ (Fig.\ref{fig4} A). The electron thus tunnels to the other layer within the first Brillouin zone, while the other processes are exponentially suppressed \cite{VolkovPRB-2023}. In the vicinity of $K_N$ the interlayer tunneling can be approximated by a constant (but see discussion in Sec. \ref{sec:subsec:topology} below): $\mathcal{H}_{tun} \approx t \sigma_1 \tau_3 $, where $\sigma_1$ is the $x$-Pauli matrix in the layer space.

We now present the general twisted Hamiltonian for TNS in the low energy limit for a circular Fermi surface, where $\epsilon({\bf K})$ is invariant under rotation and the calculations remain analytically tractable and conceptually clear. The pairs of nodes from two layers are well separated in momentum space and form isolated "valleys" (Fig. \ref{fig4}). For a single valley one has:
\begin{equation}
    \mathcal{H}_{tBSC}=v_F k_{\parallel} \tau_3 \sigma_0 + v_\Delta k_\perp \tau_1\sigma_0  -\frac{v_\Delta Q_\theta}{2}\tau_1  \sigma_3 + t \tau_3 \sigma_1.
    \label{eqn:Hamtbsc}
\end{equation}
The Hamiltonian \eqref{eqn:Hamtbsc} can be solved exactly, without having to use perturbation theory in the interlayer tunneling as is done in TBG~\cite{bistritzer2011moire}.
While the displaced Dirac cones do hybridize, no isolated moir\'e bands form (see also \cite{can2021high}).

Nonetheless, similar to TBG, at a finite but small twist angle the Dirac cone velocities $v_{F,\Delta}$ in Eq.~\eqref{eqn:Hamtbsc} are strongly renormalized. Interestingly, just as in TBG, we find this is described by a dimensionless coupling constant $\alpha =\frac{2t}{ v_{\Delta}K_N\theta} \equiv \frac{\theta_{MA}}{\theta}$  leading to
\begin{equation}
   \frac{\tilde v_{F,\Delta}}{v_{F,\Delta}} =
   \sqrt{1-\mathrm{min}\left[\alpha^2, \alpha^{-2}\right]} ; \,\,\,\, \theta_{MA}=\frac{2t}{v_{\Delta}K_N}.
   \label{eq:velocity}
\end{equation}
Despite the anisotropy $v_F\neq v_\Delta$, both velocities are renormalized in the same way and simultaneously vanish at a ``magic'' twist angle $\theta_{MA}$. In the circular Fermi surface model, 
(Fig. \ref{fig4}, B on the bottom left) the positions of two Dirac points merge at $\theta_{MA}$ in momentum space resulting in a quadratic band touching, that generates a finite density of states at the Fermi energy.

While the downward renormalization of $\tilde v_{F,\Delta}$ has been shown to occur with a non-circular Fermi surface \cite{VolkovPRB-2023} and including self-consistency effects \cite{Tummuru-Franz-PRB-2022} (see Fig.~\ref{figpd}), the exact nature of the quasiparticle dispersion at $\theta_{MA}$ depends on details: for non-circular Fermi surface a semi-Dirac point is formed (Fig. \ref{fig4}, B on the bottom right), while for non-symmetry protected nodes (such as ones that may arise in extended $s$-wave order parameter) the complete softening is avoided altogether \cite{volkov_2025}. On the other hand, in multilayer TNS with alternating or chiral twists between successive layers, an enhanced flattening of the spectrum (for a circular Fermi surface model) was reported \cite{Tummuru-Franz-2022}. In all of these cases the partial flattening (or softening) of the Dirac spectrum enhances the strength of interactions between the BdG quasiparticles that can result in secondary instabilities within the superconducting state that we review below in Sec.~\ref{subsec:correlations} (and see Fig.~\ref{figpd}, top left and bottom right).

It is instructive to consider estimates of the magic-angle for different quasi-2D nodal superconductors. For \bscco (based on ab initio results, see discussion in \ref{sec:subsec:topology} below)  $\theta_{MA}\approx 2.8^\circ$, while estimates for organic  (BETS)$_2$GaCl$_4$ and heavy fermion CeCoIn$_5$ are $1^\circ$ and $14^\circ$, respectively \cite{VolkovPRB-2023}. All of these materials are (putative) $d$-wave superconductors; in all cases $\theta_{MA}$ is very different from $\theta_{TRSB}$ resulting from order parameter hybridization, discussed in Sec. \ref{sec:OP}.

The suppressed energy scales in moir\'e materials typically lead to an enhanced tunability of the electronic bands by external perturbations. The same holds for BdG bands in TNS (summarized in  Table~\ref{tab1}), albeit the effects of external perturbations are drastically altered by the superconducting nature of the system. Each external field mentioned in Table~\ref{tab1} introduces additional terms in the Hamiltonian in Eq.~\eqref{eqn:Hamtbsc} (detailed discussion in Ref.~\cite{VolkovPRB-2023}). For example, the BdG quasiparticles can  be ``gated,'' in analogy with electronic 2D materials, by applying 
a magnetic field that results in the Zeeman coupling $\mathcal{H}_Z={\bf h}\cdot{\bf s}$ or an in-plane current (that imposes a finite Cooper pair momentum \cite{campuzano2016,zhu2021discovery} $\Delta \to \Delta e^{i{\bf Q}_{CP} {\bf r}}$) that is represented by $\mathcal{H}_{I_{\parallel}}={\bf v}_F\cdot{\bf Q}_{CP}$, both of which produce BdG quasiparticle pockets out of the Dirac cones. In addition to the Zeeman coupling,  orbital effects of the magnetic field start to play an important role \cite{tinkham2004introduction} as vortices [Abrikosov for out-of-plane field and Josephson for in-plane \cite{VolkovPRL-2023}, see Fig.~\ref{fig:gap} (Right)] develop. Their effects are related to the underlying topology in the ground state wavefunction and will be discussed in \ref{sec:subsec:topology}.

Potentially the most striking consequences, unique for TNS, are obtained when an {\it interlayer} current is considered. As discussed in Sec. \ref{sec:OP}, the interlayer current is uniquely related to the phase difference of the order parameters across the interface via Eq. \ref{eq:cur}. Conversely, driving a current through the TNS interface results in $\Delta_{1/2}\rightarrow\Delta_{1/2}e^{\pm i \varphi/2}$. We note that the same effect occurs on entering the time-reversal broken phase around $\theta_{TRSB}$ discussed in Sec. \ref{sec:OP}. As we discuss in the next section, the time reversal symmetry breaking in both cases transforms TNS into topological superconductors.

\subsection{Topology in the BdG bands}
\label{sec:subsec:topology}
As we have just mentioned, applying an interlayer current introduces a phase difference $\varphi$ between the order parameter of the two superconductors. In the continuum model, Eq. \ref{eqn:Hamtbsc}, this introduces a new term that is nonzero at the Dirac points and for $\varphi\ll1$ is: $\delta H_{\varphi\ll1} =\frac{v_\Delta Q_\theta \varphi}{2}\tau_2$. It opens a gap by introducing a Dirac mass term with the same sign for the two Dirac points in each valley (Fig.~\ref{fig4} C). 
This effect is somewhat similar to the effect of aligning a graphene layer with hBN~\cite{Bultnick-PhysRevLett.124.166601-2020}. But in remarkable contrast, one can show that the Berry curvature of all the gapped nodes is the same for TNS \cite{VolkovPRL-2023} resulting in the total Chern number $|C_{\mathrm{tot}}|=2N_{nodes}$  \cite{bernevig2013topological}. For a bilayer of \bscco, for example, where every layer contains two CuO$_2$ layers, this would lead to $|C_{\mathrm{tot}}|=8$. The topological gap opened by this mechanism, is present at all twist angles \cite{VolkovPRL-2023} and temperatures \cite{can2021high}, and can withstand  a finite amount of disorder ~\cite{VolkovPRL-2023,jxzhu_2024} (although its value may vary, see below). This is in stark contrast to other moir\'e platforms (Table \ref{tab1}), where fine tuning the twist angle and low enough temperatures (to enable the moir\'e correlation effects) are necessary to realize topological states. 

Such robustness of topology in TNS can be traced back to symmetry principles \cite{VolkovPRL-2023}; here we present the argument for a $d_{x^2-y^2}$ superconductor on a square lattice, with $D_{4h}$ point group and perturbatively small twist. Twist reduces the point group from $D_{4h}$ to $D_4$, acting as a perturbation belonging to $A_{1u}$ representation. The current behaves as the $A_{2u}$ representation (but is odd under time-reversal). Acting together allows the $d_{x^2-y^2}$ ($B_{1g}$ representation) and  $i d_{xy}$ symmetries ($B_{2g}$ representation) to mix, as $A_{1u} \times A_{2u} = B_{1g} \times B_{2g} = A_{2g}$. Superconductors with the order parameter $d_{x^2-y^2}+i d_{xy}$ are one of the well-known examples of chiral topological superconductivity~\cite{kallin2016chiral}, with $|C|=2$ for a single layer, consistent with the results above. Interestingly, a magnetic field perpendicular to the plane of TNS has exactly the same symmetry properties, as the combination of twist and current (the $A_{2g}$ representation, breaking time reversal). Indeed, an induction of $d_{x^2-y^2}+i d_{xy}$ pairing by magnetic field was proposed as early as 2000 \cite{balatsky2000}; however due to the requirement of fields below $H_{c1}$ to avoid Abrikosov vortices, this proposal is likely to yield only weak effects.

\begin{figure}[h]
\centering
\begin{minipage}{0.45\textwidth}
    \includegraphics[width=\textwidth]{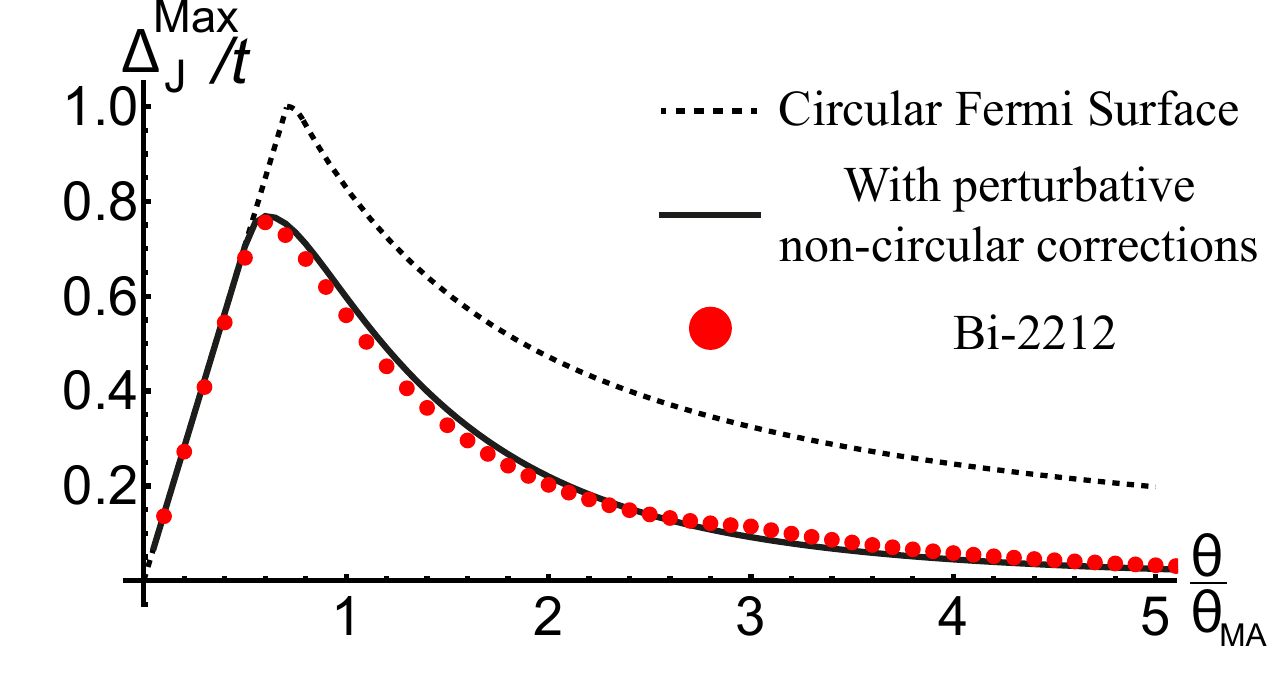}
\end{minipage}
\hfill
\begin{minipage}{0.45\textwidth}
        \includegraphics[width=0.7\textwidth]{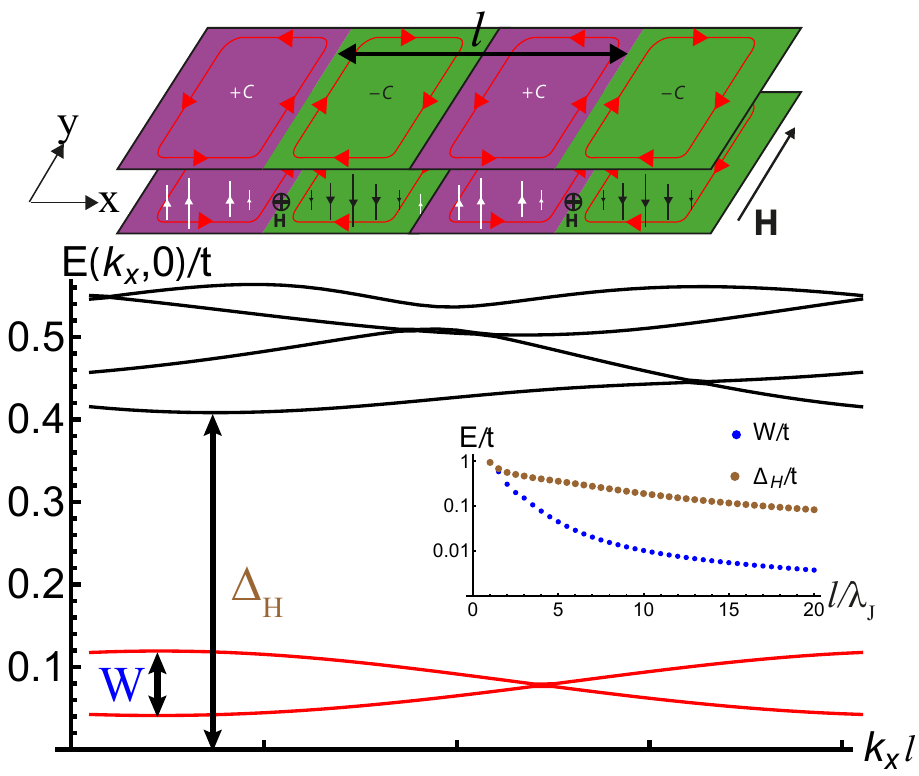}
\end{minipage}
\hfill
\caption{
(Left) The topological gap (maximized over the applied current induced phase difference $\varphi$) computed within the low energy description of TNS from Eq.~\eqref{eqn:Hamtbsc} for a circular Fermi surface, the full Fermi surface of BSCCO and with approximate non-circular corrections. The Chern number in this regime is $|C|=2N_{\mathrm{nodes}}$, Figure from Ref.~\cite{VolkovPRL-2023}. 
(Right; top) Schematic of applying  an in plane magnetic field to TNS, which induces an periodic pattern of domains with alternating Chern number (shown with green and purple) with chiral edge modes (red lines) that co-propagate at their boundary. (Right; bottom) The edge modes form an energy band with a small bandwith $W$ that is separated from the rest of the BdG modes by gap $\Delta_H$.
Figure from Ref.~\cite{VolkovPRL-2023}.
}
\label{fig:gap}
\end{figure}

We now discuss the value of the topological gap that can be induced in TNS. The maximal gap that can be induced by a phase difference in the low-twist angle continuum model is presented in Fig.~\ref{fig:gap} as a function of twist comparing results for a circular Fermi surface and a typical \bscco Fermi surface (denoted Bi-2212). One observes that the gap value reaches its maximum (which is of the order of interlayer tunneling $t$) close to $\theta_{MA}$ and is then smoothly reduced, but more rapidly for non-circular Fermi surfaces. For strong interlayer tunneling ($t$ larger than the maximum value of the superconducting order parameter) gaps of the order $t$ have been demonstrated in the vicinity of $45^\circ$ for the $d$-wave case \cite{can2021high}. Special considerations need to be taken into account when applying these results for \bscco. In that case, the single-particle tunneling between layers is dependent both on momentum and twist, due to the conducting electrons stemming from Cu $d_{x^2-y^2}$ orbitals (which is different from, and not necessarily related, to $d_{x^2-y^2}$ superconductivity). As a result, the interlayer tunneling has the functional form~\cite{Song-2022} 
\begin{equation}
    t_z({\bf K}_x,{\bf K}_y,\theta)=t_1(\cos K_x^{\theta/2} - \cos K_y^{\theta/2})(\cos K_x^{-\theta/2} - \cos K_y^{-\theta/2}) +t_0(\theta)
    \label{eqn:interlayertunneling}
\end{equation}
 \cite{Markiewicz-oneband-2005,Song-2022}, where $K_{x,y}^{\pm \theta/2}$ are components of ${\bf K}$ rotated by $\pm \theta/2$. The first term vanishes along the nodal lines of both layers, while $t_0(\theta)$ has to vanish exactly at $45^\circ$ due to a similar argument, as for order parameters in Sec. \ref{sec:OP}.
As a result, the study in Ref.~\cite{Song-2022} found a nontopological state to occur around $45^\circ$, see Fig.~\ref{figpd} (bottom left). It has been shown, however, that if the tunneling breaks momentum conservation and point group symmetries, the topological gap reappears~\cite{Haenel-2022}. We highlight that these arguments do not affect the topological gap induced by current at low twist angles $t_0(\theta\ll 45^\circ)$. Presently, the interlayer nodal tunneling has not been measured experimentally, while estimates based on ab-initio results \cite{VolkovPRL-2023, Markiewicz-oneband-2005} suggest a value of $\approx 1$ meV for the topological gap to be achievable in \bscco.

The ideas reviewed above have been extended in several directions. First, applications to triplet superconductors have been considered. Twisting  $p$-wave atomic wires at $90^{\circ}$~\cite{Tummuru-Franz-2021} has been predicted to realize the $p+ip$-like state. The application of a spin current in triplet TNS has been shown to realize $\mathbb{Z}_2$ topological superconductivity for systems with an odd (e.g. 3)  number of layers \cite{Lucht-2024}. Interfacing TNS with topological insulators has been shown to be a route to realize Majorana modes at high temperatures in vortex cores \cite{mercado2022} or at the topological insulator corners \cite{lu2023majorana}. Consequences of increased layer number for topology has been considered more broadly \cite{Tummuru-Franz-PRB-2022,Tummuru-Franz-2022,lucht2025}. An experimentally relevant setup is an interface between flakes of finite thickness, i.e. $N$-layered flakes. For such $N+M$ TNS, it has been found that while the whole flakes become topological (with Chern number growing with thickness), the gap size is reduced \cite{Tummuru-Franz-PRB-2022} as (thickness)$^{-1}$ on average \cite{lucht2025}. Remarkably, inducing the topological state via current \cite{lucht2025} allows one to drive topological transitions to states with {\it higher} Chern numbers. Finally, some of the effects arising in TNS due to topology can appear in non-nodal superconductors too: specifically, the thermal Hall effect has been predicted to occur and be a probe of the  gap anisotropy in $s$-wave superconductors, such as FeSe \cite{han2023}.

\subsection{Probes of topological superconductivity in TNS}
\label{subsec:topexp}
We now summarize the experiments that can probe topological states in TNS. From a theoretical perspective, the ``gold standard" observation of topological superconductivity is the quantization of the thermal Hall conductivity $\kappa_{xy}$ (and, for singlet TNS spin Hall effect \cite{senthil1999}).
which directly relates to the Chern number $\kappa_{xy}/T\propto C_{tot}$. Experimentally, it may be extremely challenging to confine the heat flow to within a bilayer. A detailed \cite{lucht2025} analysis of twisted multilayer flakes revealed, however, that at intermediate temperatures (larger than the topological gap), the thermal Hall conductivity is {\it independent} of flake thickness due to a cancellation between increasing number of edge modes and decreasing gap. While this does not provide a definite proof of topology, the magnitude of $\kappa_{xy}/T$ is expected to be much larger in the topological case than in the non-topological one \cite{han2023}. 

Another signature of topological superconductivity in TNS are edge states \cite{can2021high,lucht2025}. Scanning tunneling microscopy \cite{yin2021probing} measurements can detect both the bulk gap and the density of states associated with the zero-energy edge modes, which can also be probed with point- or planar-junction spectroscopy \cite{lofwander2001}. Furthermore, for a single layer twisted on top of an $N$-layer flake it has been demonstrated that the top layer would contain a spectral gap feature of almost the same size as in a bilayer \cite{lucht2025}. Finally, the edge modes can also result in an observable chiral current
\cite{franzedge} that can be detected with SQUID~\cite{RevModPhys.71.631-Koelle-1999} or NV center magnetometers~\cite{RevModPhys.92.015004}.

An issue identified above specifically for \bscco is the possibility that the topological gap near $45^\circ$ may be too small. Inducing a topological state by applying an interlayer current solves that problem, allowing one to maximize the gap near $\theta_{MA}$. Moreover, a contact-less version of this setup is possible by applying an in-plane magnetic field, see Fig.~\ref{fig:gap} (Right). In \cite{VolkovPRL-2023}, it has been shown that an in-plane magnetic field will induce alternating stripe domains with chiral edge modes flowing along the domain walls. These can be detected either using the same techniques as discussed above; in addition, the induced edge modes are expected to contribute to the usual (non-Hall) thermal conductivity, $\kappa_{xx}$.

\subsection{Correlations in the BdG bands}
\label{subsec:correlations}
The question of interquasiparticle interactions in superconductors needs clarifying: indeed, the order parameter itself arises due to the interactions. However, as long as interlayer tunneling is much weaker than the order parameter magnitude $\Delta_0$ (and temperature is much smaller than $T_c$), the ``back-action" of the quasiparticle reconstruction by twist on the order parameter is suppressed \cite{VolkovPRB-2023} at low twist angles. In that case, the flattening of the Dirac dispersion close to $\theta_{MA}$ (Fig. \ref{fig4} B) favors a secondary superconducting instability at $T^* \ll T_c$ that breaks time reversal symmetry~\cite{VolkovPRB-2023}, see Fig.~\ref{fig:gap} (Right). The nature of the secondary order reflects the subleading superconducting instability in a single flake. E.g. if the leading instability is $d$-wave and subleading (not realized in single flake) is $p$-wave, a $d+ip$ phase will emerge. Note that a multitude of competing superconducting states may occur in, e.g., the Hubbard model \cite{deng2015emergent}. The new phase occupies a ``dome" in the $T-\theta$ phase diagram with maximal $T^*$ reached at $\theta=\theta_{MA}$, see Fig.~\ref{figpd} (Bottom Right). Fully self-consistent (i.e. including the back-action of quasiparticles on the order parameter) calculations for a model with nearest-neighbor attraction \cite{Tummuru-Franz-PRB-2022} find such a transition to a $d+is$ state, extending to large twist angles until it is superseded by the $d+id$ state at large twist angles, see Fig.~\ref{figpd} (Top Left). 

Interaction effects have been shown also to impact topology close to $45^\circ$. Self-consistent modeling of the superconducting order with nearest-neighbor interaction \cite{can2021high} have shown the possibility to realize time-reversal broken phases with different Chern number ($|C|=0,2,4$) in a $d$-wave TNS (Fig. \ref{figpd}, top right).

Another focus of the field has been considering the impact of strong correlations in individual layers, and specifically in \bscco. Within a parton mean field approach to describe twisted $t-J$ bilayers~\cite{Song-2022}, a different sequence of topological phases near $45^\circ$ was predicted, taking the momentum- and twist- dependent tunneling into account as shown in Fig.~\ref{figpd} (Bottom Left). Other works focused on the time-reversal breaking instability near $45^\circ$ in the framework of a $t-J-U$ model~\cite{spalek2023}, variational cluster approximations~\cite{senechal_2022,senechal_2024,senechal_2024_2}, and cluster dynamical mean field theory including effects of varying doping and the strong correlations in each monolayer. There are also extensions to the insulating limit of undoped BSCCO and twisted Mott insulators~\cite{zhang_site_selective_2024}, which offer a new direction for twisted quantum magnets. Incidentally, ways of realizing square lattice Hubbard models (believed to be relevant for cuprates) with widely tunable parameters using moir\'e materials have been proposed \cite{eugenio2024}.

\begin{summary}[SUMMARY POINTS: Quasiparticle hybridization in TNS]
\begin{enumerate}
\item Twist reduces the Dirac velocity of quasiparticles in TNS in the vicinity of a ``magic angle" $\theta_{MA}$ which is typically much smaller than $\theta_{TRSB}$. This stipulates time-reversal breaking transitions (distinct from the one at $\theta_{TRSB}$); however, no isolated flat bands form.
\item Breaking of time-reversal symmetry (either spontaneous near $\theta_{TRSB}$ or induced by interlayer current at any twist angle) induces topological superconductivity in TNS. For weak hybridization, the topological gap value is maximal close to $\theta_{MA}$ but is nonzero at all $\theta$.
\item Around $\theta_{TRSB}$, the form of tunneling, strong correlation, and self-consistency effects modify the nature of time-reversal symmetry breaking and the topological properties; many phases survive but their positions in the phase diagrams change.
\end{enumerate}
\end{summary}

\section{Summary and Outlook}
\label{sec:conclusions}
Over the last few years, twisted nodal superconductors have emerged as a fully fledged new addition to the family of moir\'e materials, with unique challenges and opportunities. Their main distinguishing feature compared to other moire platforms (see Table \ref{tab1} for a detailed comparison) is the presence of a superconducting order parameters in separate layers, which leads to a different set of physical properties and distinct ways to manipulate these systems. The order parameter can itself be affected by twist and lead to emergent phenomena, such as time-reversal symmetry breaking near $\theta_{TRSB}$, reviewed in Sec. \ref{sec:OP}. The quasiparticles in TNS show high tunability, including magic-angle physics similar to TBG near $\theta_{MA}\neq \theta_{TRSB}$, and allow for the creation of  topological phases, both spontaneously and on demand (Sec. \ref{sec:WF}). However, the flattened Dirac quasiparticles in TNS have not been shown to form isolated flat bands; finding ways to do so may open TNS to the frontier of strongly interacting topological states~\cite{park2023observation,zeng2023thermodynamic,kang2024evidence}
and fractionalized superconductivity.

TNS experiments are at present focused on the high-temperature superconducting cuprates, chiefly \bscco. Remarkable recent progress  in fabrication of few-layer thick films \cite{yu2019,Zhao-Kim-2019} and interfaces  between 10-100 nm flakes~\cite{zhao2023,xue_2023_OP,martini2023twisted,lee2021twisted} have opened the door to investigate the physics of TNS interfaces via transport measurements. These brought about the observation of the signatures of time reversal breaking and the second-order Josephson effect \cite{zhao2023}. However, the experiments also hint at a potential interplay between different sources of time-reversal breaking and interface structure, see Sec. \ref{sec:OP}. A concerted effort for a deeper experimental understanding of the current-phase relationship for tBSCCO near 45$^\circ$, mapping out the first and second harmonic contributions across the phase diagram, may be able to settle these questions. Equally important is developing other means to accurately probe the atomic structure and the nature of superconductivity at the interface. Present proposals include optical \cite{can_pke,xiao2024optically,kaplan2025quantum}, transport \cite{shokri2024evolution,Zhao-Kim-2019} and magnetic \cite{franzedge,jxzhu_2024}
probes. Interpretation of these experiments for \bscco will require assessing the influence of strong correlation effects\cite{Song-2022,senechal_2022,senechal_2024,senechal_2024_2} and material specifics \cite{Song-2022,yuan2023,volkov_2025} on their outcomes. These developments will bring about new ways to probe correlated states and superconductivity in cuprates as well as other unconventional and 2D superconductors, already inspiring some proposals \cite{xiaoberg_2023,yuan2024phase}.

An important future direction is the development of alternative material platforms for TNS. In principle, there is no lack of candidate nodal superconductors \cite{stewart2017unconventional}, with many being highly two-dimensional. To name a few, organic (e.g., \cite{clark2010superconductivity}, heavy fermions (where epitaxial layer-by-layer growth has been established [see \cite{asaba2024evidence} and references therein], the recent discovery of van der Waals CeSiI~\cite{posey2024two}), Fe-based superconductors where epitaxial monolayer growth has been achieved,
and (possibly) Sr$_2$RuO$_4$ \cite{maeno2024} are all in that category. The realization of twisted junctions of flakes of NbSe$_2$\cite{yabuki2016supercurrent,farrar2021superconducting} and NbS$_2$\cite{zhao2022josephson} also suggests the possibility that creating twisted monolayers of TMD superconductors is within reach. While in bulk form many of these materials are believed to be conventional, there is some evidence of unconventional (in NbSe$_2$) \cite{hamill2021two} and nodal (in TaS$_2$) \cite{vavno2023evidence} superconductivity in monolayers, which also generically have singlet-triplet mixing due to the absence of an inversion center. Last, some of the 2D and moir\'e materials presently explored may be nodal superconductors, leading to a possibility of twisting moir\'e superconductors (e.g. such as superconducting TBG twisted  on superconducting TBG).

The present experiments do not directly address the question of topology or quasiparticle reconstruction near $\theta_{MA}$. Theoretically, a number of probes have been proposed to detect the topology of TNS in Sec.~\ref{subsec:topexp}. Developing new experimental platforms, such as the quantum twisting microscope \cite{inbar2023quantum} and non-linear optics~\cite{kaplan2025quantum}, may lead to a breakthrough in probing the quasiparticle reconstruction and topology.

Finally, beyond fundamental research, applications of TNS in quantum information problems are being actively developed. Several qubit designs have been proposed recently including the ``flowermon'' qubit for twisted $d$-wave bilayers at $45^{\circ}$~\cite{Brosco-Poccia-Vool-PhysRevLett.132.017003-2024}, and the ``$d$-mon'' qubit~\cite{Patel-Franz-PhysRevLett.132.017002-2024} by interfacing a $d$-wave and $s$-wave superconductor. In both cases, the second harmonic current-phase relation leads to conserved Cooper-pair parity, beneficial for coherence times and there are mechanisms for suppressing the quasiparticle effects. TNS-based heterostructures have been also shown to host Majorana modes at high temperatures \cite{mercado2022,lu2023majorana}. While several hurdles remain~\cite{confalone2025challenges}, progress to interface such a device has also been achieved through incorporating tBSCCO devices into printed circuit boards~\cite{saggau-Poccia-2023} and coupling them to a hybrid superconducting microwave resonator~\cite{velluire2023hybrid}.

Starting around 2021 in its present form\cite{vishwanath2021proposals,glazman2022jc, can2021high}, the TNS field has seen a rapid pace of development of experimental techniques, theoretical ideas, and potential applications. The ideas and methods that arose in the TNS community spread to other fields, while offering a scope for expansion to a large amount of potential platforms and setups. We hope that this review will further stimulate this process.

\begin{issues}[FUTURE ISSUES]
\begin{enumerate}
\item Detection of topological states in TNS is the most pressing challenge. Exploring and developing new probes (transport, optical, tunneling, magnetic) and experimental settings (such as using current or magnetic field control), Sec. \ref{subsec:topexp}, may provide a path forward.

\item Strong correlations of quasiparticles have not yet been demonstrated in TNS - both experimental studies near $\theta_{MA}$ and theoretical proposals to create flat quasiparticle bands need to be developed.

\item The experimental platforms for TNS can be expanded beyond cuprates to utilize the large number of candidate quasi-2D nodal superconductors including heavy fermions, organics, Fe-based superconductors,  rhombhohedral graphene, and moir\'e superconductors.

\item Tunability and high operating temperatures can make TNS attractive for quantum information applications, but the challenges for their integration into quantum devices need to be surmounted first.


\end{enumerate}
\end{issues}






\section*{DISCLOSURE STATEMENT}
The authors are not aware of any affiliations, memberships, funding, or financial holdings that
might be perceived as affecting the objectivity of this review. 

\section*{ACKNOWLEDGMENTS}
We thank our collaborators on these topics Marcel Franz, Philip Kim, Kevin Lucht, Nicola Poccia, Justin Wilson, and Frank Zhao as well as Eva Andrei, Piers Coleman, Daniele Guerci, Gabi Kotliar,  and Andy Millis for fruitful discussions. We thank Lucy Reading-Ikkanda/Simons Foundation for the help with figure design and execution. We would like to specially thank Yejin Lee, Gil-Ho Lee, Hu-Jong Lee, Jongyun Lee, Qiang Li, Richard A. Klemm, Yoshihiko Takano, Ding Zhang, Yuying Zhu and Qi-Kun Xue for sharing the data from their works. This work has been supported in part by the NSF CAREER Grant No.~DMR 1941569 and a Sloan Research Fellowship (J.H.P.).
This work was  performed  in part at the Aspen Center for Physics, which is supported by National Science Foundation grant PHY-2210452.
This research was supported in part by grant NSF PHY-2309135 to the Kavli Institute for Theoretical Physics (KITP).
The Flatiron Institute is a division of the Simons Foundation.

%





\bibliographystyle{ar-style4}
\bibliography{refs}
















\end{document}